# The Application of Coherent Microwave Scattering and Multiphoton Ionization for Diagnostics of Electric Propulsion Systems

**Adam R. Patel, Sashin L. B. Karunarathne, Nicholas Babusis, and Alexey Shashurin**

School of Aeronautics and Astronautics, Purdue University, West Lafayette, IN, USA

## Abstract

Nonintrusive measurements of plasma properties are essential to evaluate, and numerically simulate, the in-flight performance of electric propulsion systems. As a logical first step in the development of new diagnostic techniques, this work depicts the implementation of multiphoton ionization and coherent microwave scattering (MPI-CMS) in a gridded-ion accelerator operating on rare gases. Presented studies primarily comprise photoionization spectroscopy of ground & excited state-populations of both neutrals and ions – supplemented by optical emission spectroscopy and Langmuir probe derived plume properties. Results suggest the potential of MPI-CMS for non-intrusive measurements of specie number densities.

# Introduction

Ion thrusters are electrostatic, high-specific impulse rarefied propulsion systems deployed in the context of astronautics. Standard inert gas-fed embodiments typically consume 1-7 kW of power and exhibit very low thrust (~25-250 mN) but maintain excellent exhaust velocities on the order of ~ 20-50 km/s [1]. Initially debuting on the *SERT-1* [2], ion engines have found use in applications ranging from geosynchronous Earth-orbit communication satellite upkeep (XIPS thruster [3],…) to asteroid probing (*Deep Space 1* [4], *Dawn* [5], Hayabusa / Hayabusa 2 [6], …). Current state-of-the-art concerns application of the 6.9 kW NEXT-C in NASA's double asteroid redirection test (*DART*) through spacecraft impact-induced deflection of the minor-planet moon Dimorphos [7]. Aside, ion beams have found general use in a diverse array of applications – ranging from etching & ultra-precision machining (encompassing semiconductor manufacturing) [8]–[12] to focused deposition [13], scanning electron microscopes [14], helium ion microscopes [15], radiobiology [16], and particle therapy [17].

In comparison to alternate electric propulsion technological classifications, ion thrusters exhibit relatively straightforward operational principles. These devices are delegated three primary tasks: ion generation, ion acceleration, and plume neutralization [1]. In a standard thruster configuration, neutral rare-gas propellant is fed into a crossed-field electron collision-based ionization chamber. Generated ions in the resulting quasi-neutral plasma are sequentially accelerated via a biased-grid through a space-charge limited channel. To restore quasi-neutrality in the exhaust plume (and thus, alleviate spacecraft charge buildup), an electron source neutralizer is used. A decelerator grid is further utilized to reduce erosion of the accelerator grid from ion-impact sputtering [18].

Although the physics of ion thrusters are relatively well-established [19], the overarching domain of electric propulsion encompasses unresolved phenomena and topics. E.g., deriving factors which foster differences between in-space and facility performance [20], minimizing erosion rates [21]–[24], studying anomalous electron diffusion, microturbulence, axial ion-ion streaming instabilities [25]–[29], etc. The resolution of this phenomena is particularly important in the improvement of numerical simulations – seeking to partially supplant timely, expensive test and qualification campaigns. Thus, there is a need to develop spatially and temporally-resolved diagnostics for ground testing of parameters/performance of electric propulsion devices.

A variety of intrusive techniques have been historically implemented for broad plasma characterization, but these diagnostics are largely perturbative and often suffer from poor spatial and temporal resolutions [30]. Prior art includes Langmuir probes (plasma potentials, electron densities, electron temperatures), Faraday probes (local ion charge fluxes, local electron charge fluxes), retarding potential analyzers (ion energy distribution functions), and Wien ExB filters (charge-states, ion energy distribution functions) [31]–[34]. Limitations of invasive techniques quickly sparked interest in nonintrusive / passive diagnostics – including optical emission spectroscopy (populations corresponding to specie state transitions), laser-induced fluorescence (populations corresponding to specie state absorption then emission, ion drift velocities), and laser Thomson scattering (electron number densities, temperatures, and distribution functions) [30], [35], [36]. However, these non-intrusive methods exhibit their own set of disadvantages – typically referring to time and path-averaging, sensitivity-issues, a difficulty in absolute calibration, and / or great experimental complexities. To further broaden the diagnostic options pool and (potentially) circumvent pre-existing technique drawbacks, multiphoton ionization + coherent microwave scattering (MPI-CMS) may suffice as an appropriate novel diagnostic for non-intrusive, single-shot standoff measurements of electron propulsion plasma parameters.

Coherent microwave scattering (CMS) is a technique used to measure the total number of electrons ($N_e$) in an unmagnetized, classical small plasma object through the constructive elastic scattering of microwaves [37]. Specifically, the diagnostic is used when the microwave skin depth $\delta \gg$ plasma diameter $D$ and microwave wavelength $\lambda \gg D$, plasma length $L$. In this case, the plasma is periodically polarized and emits short-dipole-like radiation which can be attributed to $N_e$ through a calibration sample [38]. Note that this condition is not satisfied by the ion thruster plume but, rather, an electron number density inhomogeneity induced by the MPI process – to be further elaborated in the results. From spatial phase related considerations, wavelet contributions from the larger plasma body will not interfere constructively and, correspondingly, produce a negligible signal.

CMS has found use in applications ranging from electron rate measurements [38]–[45] to trace species detection [46], gaseous mixture and reaction characterization [47]–[49], molecular spectroscopy [50], and standoff measurement of local vector magnetic fields in gases through magnetically-induced depolarization [51]. Advantages of the technique include a high sensitivity (total number of electrons detectable $> 10^7$), good temporal resolution, single-shot acquisition, and the capability of time gating due to continuous scanning. Often, as in this study, CMS is coupled with laser-induced resonance-enhanced multiphoton ionization (REMPI) – a special case of multiphoton ionization (MPI) which provides an efficient pathway for species' state-selective ionization, following the m+n (m simultaneous photons to an excited intermediate state and n subsequent photons to continuum) ionization scheme. Further, the nonlinear nature of resonant photoionization adds measurement locality at the focal point of the laser. REMPI is highly selective and can ionize trace species, can be used to perform detailed spectroscopy (including fine structure populations, selectivity rules, spin polarizations [52]), measure temperature [50], and is directly linked to various phenomena (such as Doppler shifting / broadening, homogeneous collisional broadening, the Stark effect, and the Zeeman effect). Compared with CMS alone, the inclusion of REMPI allows selective species ionization – enabling (with proper calibration) a method to diagnose the number density of the selected species / state [47].

Recent extension of CMS to the Thomson low-pressure regime [53] enables application of the technique to electric propulsion systems. The Thomson scattering regime refers to the instance where the electron mobility (when driven by a microwave field) is unbound by collisions (Shneider-Miles scattering) and polarization effects (Rayleigh scattering) – accomplished when the microwave generator frequency greatly exceeds the effective collisional frequency and modified plasma frequency. In this case, a detailed knowledge of plasma parameters is not required to characterize scattering and, even further, the Thomson regime shares a space with regions of low optical nonlinearities (from reduced background number densities). Thus, MPI-CMS (or the subset REMPI-CMS) may suffice as an appropriate diagnostic for non-intrusive, single-shot standoff measurements of specie / state populations in EP systems – with the potential for further use in applications ranging from vector magnetic field mapping to velocity measurements via Doppler-shifted photoionization spectroscopy.

As a natural first step in assessing the applicability of REMPI-CMS in electric propulsion, this study specifically investigates integration of the diagnostic in an ion accelerator operating on noble gas propellants. The continuous, relatively-simple (no **B**-field) operation of ion accelerators, alongside microwave transparency due to sufficiently-low plume electron number densities, advocate the devices as an ideal demonstration candidate.

# Methodology

Schematics of the experimental assembly are depicted in **Figure 1** and **Figure 2**.

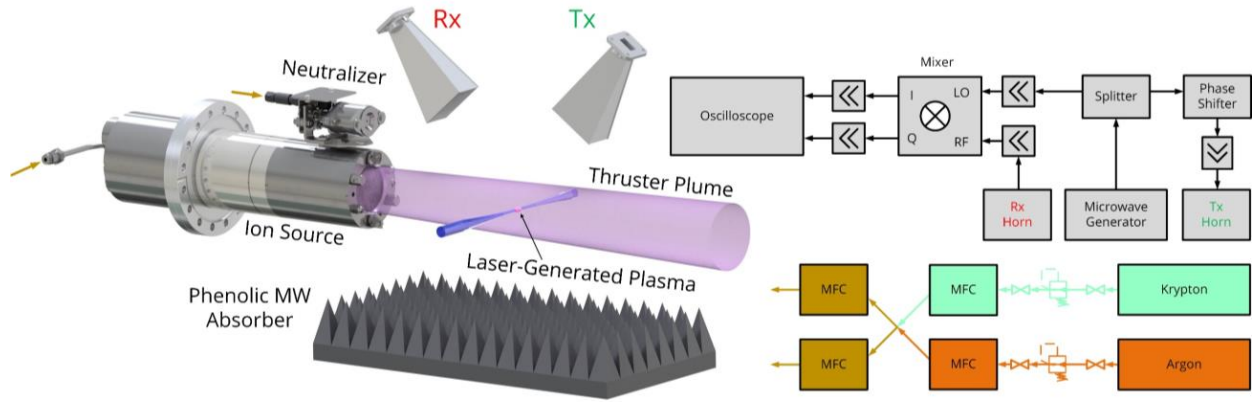

**Figure 1**: Experimental configuration for the implementation of MPI-CMS in an ion accelerator.

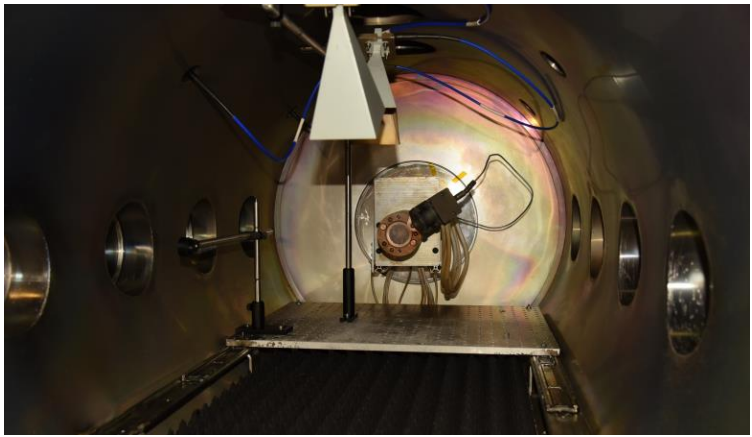

**Figure 2**: Photograph of the experimental assembly.

A full description of the vacuum chamber, nanosecond laser system, and microwave scattering circuitry can be found in [53]. Experiments were conducted in a diffusion-pumped, ionization gauge-equipped vacuum chamber lined with a phenolic-based high power microwave absorber. 4.93 ± 0.25 ns FWHM, linearly polarized laser pulses from 210 – 400 nm (FWHM 0.1 nm) were delivered by a 10 Hz, broadly tunable Ekspla NT342 Q-switched Nd:YAG laser system. The 7 mm diameter top-hat beam with full divergence angle < 2 mrad was focused through a f = 175 mm plano-convex spherical lens to a peak laser intensity on the order of ~ $10^9$ W/cm$^2$ for plasma generation. From our previous work [53], the case of (2+1) REMPI (~ cubic laser intensity process) generates a plasma object that can be regarded as an oblate spheroid with upper bound semi-axis estimates of $\mathfrak{a} = \mathfrak{b} = 280$ μm and $\mathfrak{c} = 4.5$ mm – our effective measurement spatial resolution. The ellipsoid is further reduced in volume for a (3+2) process. Following this trend, the case of non-resonant one photon ionization will produce an electron number density distribution consistent with the laser's Gaussian profile.

For CMS measurements, the microwave field scattered off the plasma volume was correspondingly measured by a receiving (Rx) horn coupled with an in-phase / quadrature (I/Q) mixer-based homodyne detection system that provides an output voltage $V_S$ proportional to the scattered electric field. First, an Anritsu 68369B synthesized signal generator was used to produce a continuous microwave signal at 11 GHz (FWHM 1 MHz). The microwaves were then amplified and split. One arm was connected to an I/Q mixer local oscillator (LO) port, with a confirmed operational power in the linear mixer regime (10 – 13 dBm). The other arm was then further amplified, isolated, and sent to a pyramidal 20 dB transmitting (Tx) horn for plasma irradiation. A receiving horn was then installed and connected to the I/Q mixer radio frequency (RF) port to detect the scattered electric field (proportional to the number of electrons). The in-phase and quadrature channels of the mixer were subsequently connected to a WavePro 735Zi 3.5 GHz oscilloscope with 50 Ohm DC termination. Calibrated amplification of the MW scattering signals (before the RF port) and the I/Q channels were used as needed to improve system sensitivity. Note that absolute calibration of in-phase CMS via a dielectric scatterer (attribution to the total number of electrons generated) is not included in this work but is an available feature of CMS.

The gridded ion source considered is a flange-mounted KDC-40 ion accelerator with complementary LFN neutralizer. Operation of the device is governed by a set of four controllers and functions in the regime where beam current does not exceed the limit which will cause direct impingement of energetic ions on the accelerator grid. Further, the accelerator voltage is greater than the electron backstreaming limit to avoid false contribution to the indicated ion beam current. Two noble gas propellants (argon, krypton) were respectively fed into the combined neutralizer-ion source system via a mass flow controller.

A test matrix for relevant operational parameters of the gridded-ion accelerator is shown in **Table 1**. Considered beam currents $I_B$ (fixed for a given voltage) are well-under the fundamental Child-Langmuir limit. As a simple estimate, the relative electron / ion number densities ($n_e \approx n_i$) between various $V_B$ can be evaluated:

$$I_B \propto n_e\sqrt{V_B} \qquad \text{Eq. 1}$$

**Table 1**: Operational parameters of the KDC-40 gridded-ion accelerator.

| Standard Anode Flow Rate (SCCM) | Fixed Neutralizer Flow Rate (SCCM) | Noble Gas | Standard Background Pressure (Torr) | Beam Voltage $V_B$ (V) | Beam Current $I_B$ (mA) |
|---|---|---|---|---|---|
| 10 | 6 | Argon | $3.6 \cdot 10^{-4}$ | 400 | 14 |
| | | | | 600 | 30 |
| | | | | 800 | 46 |
| | | Krypton | $5 \cdot 10^{-4}$ | 400 | 14 |
| | | | | 600 | 30 |
| | | | | 800 | 46 |

Note that the magnitude of LFN neutralizer current is equivalent to the beam current. Aside, a neutralizer is redundant in the considered experimental setup due to the KDC-40 grounding.

To supplement REMPI-CMS results, 1.) optical emission spectroscopy (OES) and 2.) Langmuir probe measurements were conducted at the laser focal position (~ 500 mm from the accelerator). 1.) A calibrated Ocean Optics USB4000, with the range 300-900 nm & focused via a NIKKOR 24-120 mm f/4 lens, was used for OES-based general identification of excited species. 2.) Langmuir probes were used to extract electron temperatures and number densities. These values were used to verify both microwave transparency and relative ion-number density measurements from resonance-enhanced multiphoton ionization of ions. A transformer circuit was installed to rapidly determine the V-I dependency through an alternating bias, and the Langmuir probe current was measured through a 20 kΩ shunt resistor. Plasma density and electron temperature were correspondingly determined from the experimentally measured V-I characteristics, represented by a Boltzmann relation with Maxwellian velocity distribution (prior to electron saturation). In experiment, the copper disk electrode was oriented parallel to the plume flow. This situation refers to the collection of Bohm's saturation current and mitigates the influence of directed ion velocity.

## Results and Discussion

We initially evaluate viability of the proposed environment before continuing forward with the implementation of REMPI-CMS in a gridded-ion accelerator. Namely, providing quantitative evidence for sufficiently-detectable photoionization at rarefied pressures (with the considered nanosecond laser features) and microwave transparency of the exhaust plume. First, in a krypton fed configuration, the scattering waveform for (2+1) 212.5 nm resonance-enhanced multiphoton ionization Kr $4p^6(^1S_0) \underset{2ph}{\rightarrow}$ Kr $4p^5(^2P^{\circ}_{3/2})5p\ ^2[1/2]_0 \underset{1ph}{\rightarrow}$ Kr$^+$ $4p^5\left(^2P^{\circ}_{3/2\ \underline{or}\ 1/2}\right)$ (with resulting photoelectron energies of $\mathcal{E}_{pe} \approx 3.5$ eV $\underline{or}$ 2.8 eV, respectively, from multiphoton excess estimates) was recorded at pressures $P = 3.6 \cdot 10^{-4}, 4.85 \cdot 10^{-4}$ Torr without the presence of plasma – suggesting adequate detector sensitivity in **Figure 3** ($N_e \propto V_S \propto$ the scattered electric field).

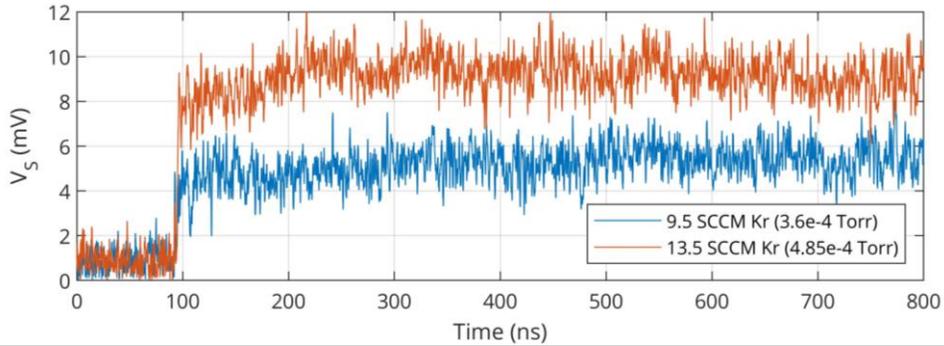

**Figure 3**: Detection of (2+1) 212.5 nm REMPI of krypton at rarefied pressures.

Next, electron number densities and temperatures were derived from Langmuir-probe V-I curves – shown in **Figure 4**. These results are summarized in **Table 2** below. In the most extreme instance of the electron number density $3.3 \cdot 10^{16}$ m$^{-3}$ (an ionization degree of $\alpha \approx 0.005$), the collisionless permittivity of the plasma can be estimated for 11 GHz microwaves: $\varepsilon = 1 - \frac{\omega_p^2}{\omega^2} = 0.978 > 0$, well above the transition to evanescence / reflection. The considered ion thruster background can thus be regarded as CMS-transparent. Furthermore, contribution of the plasma plume to the microwave scattering signal V$_S$ was confirmed to be negligible (without the laser) – indicative of destructive and/or incoherent scattering. Physically, this destructive scattering can occur for phase-locked electrons exhibiting uniform and continuous relative scattered phase distributions (at the detector) on the interval $[0, 2\pi)$ – enabled by large plasma dimensions (the ion thruster plume) relative to the microwave wavelength. Withholding a non-uniform plume electron number density distribution, phase randomization from thermal motion, microscopic density fluctuations, and Doppler broadening will produce some small scattered power proportional to the total number of electrons $N_e$. However, a relatively localized dense plasma ellipsoid (produced by the laser) can produce constructive wavelet contributions with scattered power dependence $N_e^2$ – far exceeding the plasma plume contribution [53].

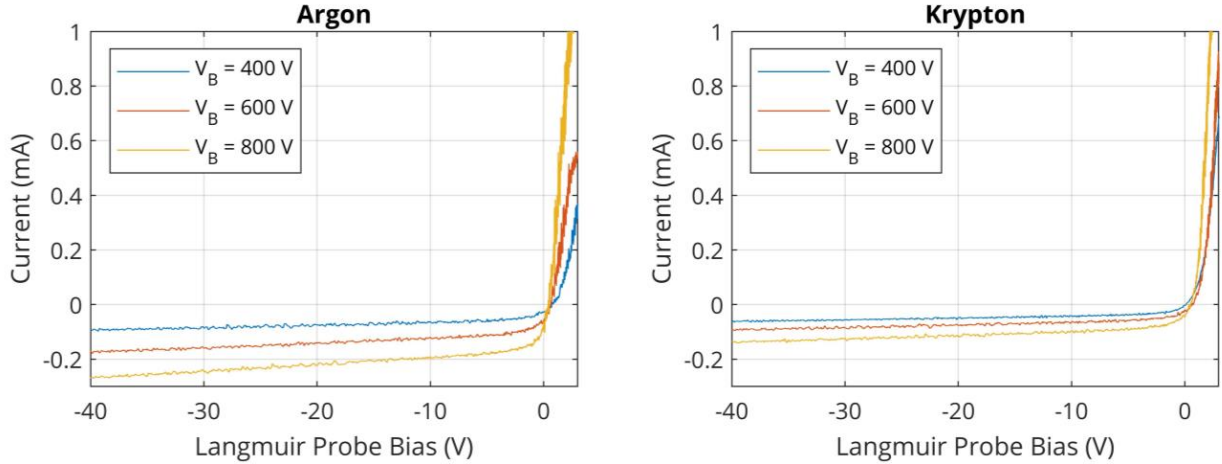

**Figure 4**: Langmuir probe VI curves for various ion thruster operational parameters.

**Table 2**: Langmuir-probe derived electron number densities and temperatures.

|  | **Argon** | | | **Krypton** | | |
|---|---|---|---|---|---|---|
| **$V_B$ (V)** | 400 | 600 | 800 | 400 | 600 | 800 |
| **$n_e$ (m$^{-3}$)** | 1.0e16 | 1.8e16 | 3.3e16 | 1.1e16 | 1.7e16 | 2.6e16 |
| **$T_e$ (K)** | 23500 | 20000 | 17500 | 14700 | 13600 | 11500 |

Compatibility of the considered background with REMPI-CMS permits microwave-based detection of photoionization embedded inside the ion engine plume. The resulting photoionization spectroscopy will reflect underlying exhaust constituents: predominantly composed of ground-state neutral atoms, excited-state neutral atoms, and singly-ionized species. For a qualitative characterization of these populations, the presence of unique quantum state transitions (strictly electronic for monatomic species) can be measured through optical emission spectroscopy – as depicted in **Figure 5**.

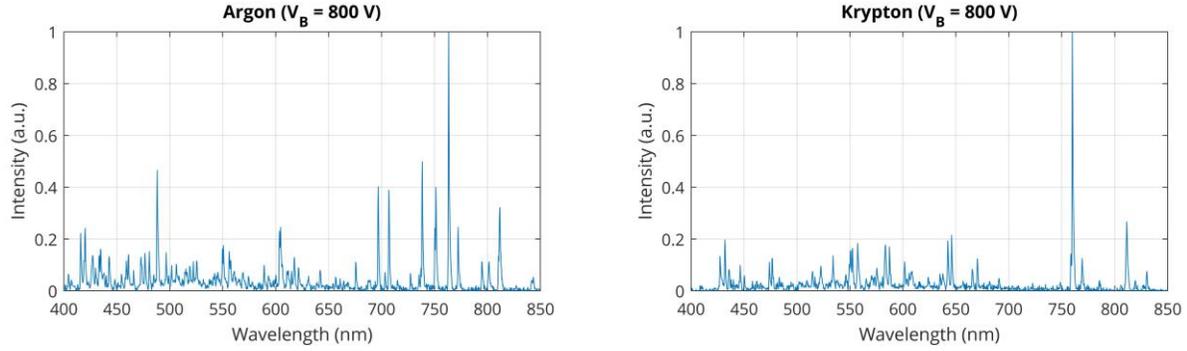

**Figure 5**: Calibrated optical emission spectroscopy of the ion thruster plume.

Identification of various species / transitions are depicted in **Table 3**. A complete analysis of observed lines is excessive and beyond the spectrometer resolution – see [54] for a full survey on transitions (particularly, those with low Einstein coefficients $A_{ki}$ or in ions). However, just a few entries in **Table 3** suggest significant excited species near the ionization threshold which can be readily ionized by single UV photons. These contributions must be considered to isolate the effect of REMPI for relative population measurements. Furthermore, the considered plasma may contain long-lifetime (~ 50 s) metastable states undetectable by emission [55], [56].

**Table 3**: A few relevant transitions in argon and krypton [54].

| Ion | $\lambda$ (nm) | Rel. Int. | $A_{ki}$ (s$^{-1}$) | Lower Level | Upper Level |
|---|---|---|---|---|---|
| Ar II | 488.0 | 0.47 | 8.23e7 | 3s$^2$3p$^4$($^3$P)4s $^2$P 3/2<br>138243.6 cm$^{-1}$<br>17.14 eV | 3s$^2$3p$^4$(3P)4p $^2$D° 5/2<br>158730.3 cm$^{-1}$<br>19.68 eV |
| Ar I | 738.4 | 0.50 | 8.50e6 | 3s$^2$3p$^5$($^2$P°$_{3/2}$)4s $^2$[3/2]° 1<br>93750.6 cm$^{-1}$<br>11.62 eV | 3s$^2$3p$^5$($^2$P°$_{1/2}$)4p $^2$[3/2] 2<br>107289.7 cm$^{-1}$<br>13.3 eV |
| Ar I | 763.5 | 1.00 | 2.45e7 | 3s$^2$3p$^5$($^2$P°$_{3/2}$)4s $^2$[3/2]° 2<br>93143.8 cm$^{-1}$<br>11.55 eV | 3s$^2$3p$^5$($^2$P°$_{3/2}$)4p $^2$[3/2] 2<br>106237.6 cm$^{-1}$<br>13.17 eV |
| Kr I | 645.6 | 0.22 | 6.65e6 | 4s$^2$4p$^5$($^2$P°$_{3/2}$)5p $^2$[5/2] 3<br>92294.4 cm$^{-1}$<br>11.44 eV | 4s$^2$4p$^5$($^2$P°$_{3/2}$)6d $^2$[7/2]° 4<br>107778.9 cm$^{-1}$<br>13.36 eV |
| Kr I | 760.2 | 1.00 | 2.73e7 | 4s$^2$4p$^5$($^2$P°$_{3/2}$)5s $^2$[3/2]° 2<br>79971.7 cm$^{-1}$<br>9.92 eV | 4s$^2$4p$^5$($^2$P°$_{3/2}$)5p $^2$[3/2] 2<br>93123.3 cm$^{-1}$<br>11.55 eV |
| Kr I | 811.3 | 0.27 | 3.61e7 | 4s$^2$4p$^5$($^2$P°$_{3/2}$)5s $^2$[3/2]° 2<br>79971.7 cm$^{-1}$<br>9.92 eV | 4s$^2$4p$^5$($^2$P°$_{3/2}$)5p $^2$[5/2] 3<br>92294.4 cm$^{-1}$<br>11.44 eV |

Waveforms for coherent microwave scattering off plasma-embedded photoionized filaments are depicted in **Figure 6** for various laser wavelengths and ion engine propellants ((a)-

(c) krypton, (d) argon). Subfigure (a) corresponds to the previously-discussed 212.5 nm krypton (2+1) REMPI mechanism. Observed scattering signals between considered ion beam accelerator voltages are qualitatively similar and suggest minimal background perturbation (see [52] for instances of laser-induced avalanche ionization – numerical simulations are likely required to confirm non-invasive characteristics of the nanosecond pulse). Temporal decay in the measured intensity can be attributed to a deviation in constructive scattering from significant electron diffusion. As a general trend, $V_S$ for $\lambda_L = 212.5$ nm appears to be proportionally correlated with the ion beam voltage $V_B$. This dependency seems counterintuitive – given an accompanying increase in the plume electron number density at sufficiently low ionization degrees such that $V_S$ should remain nearly invariant. However, the presence of excited species enables an avenue for single-photon ionization with a significantly greater PI rate (omitting the corresponding number density term) than REMPI. Even further, one-photon photoionization scaling laws enable a larger plasma volume compared to the case of (2+1) REMPI (which would appear as a dense ellipsoid core in the larger plasma structure). Contribution from ponderomotive and quiver-collision induced ionization is assumed negligible for the rarefied pressures and photon frequencies considered.

The influence of single-photon ionization can be visualized in **Figure 6**(b) for an off-resonant laser wavelength of 215 nm. This signal can be effectively subtracted from $V_S$ for 212.5 nm to isolate the effect of REMPI. Aside, exclusive ionization of excited species may be possible through optical and/or infrared multiphoton absorption to an intermediate energy level (subsequently followed through to continuum). This may also yield a novel way of detecting metastable states, which conventionally evade optical emission spectroscopy.

Inclusion of krypton-based plasma enables a new resonant-photoionization mechanism near $\lambda_L = 214$ nm. This process corresponds to the (3+2): $\text{Kr}^+ 4p^5\ ^2P^\circ_{3/2} \xrightarrow[3\text{ph}]{} \text{Kr}^+\ 4p^4(^3P)5p\ ^4D^\circ_{1/2} \xrightarrow[2\text{ph}]{} \text{Kr}^{+2}$ Or $\text{Kr}^+ 4p^5\ ^2P^\circ_{3/2} \xrightarrow[3\text{ph}]{} \text{Kr}^+\ 4p^4(^3P)5p\ ^2P^\circ_{3/2} \xrightarrow[2\text{ph}]{} \text{Kr}^{+2}$ in singly-ionized krypton with $\mathcal{E}_{pe} \approx 4.6$ eV, where upper **Figure 6**(c) indicates $V_S \propto V_B \propto n_e$ unexplainable solely by **Figure 6**(b). Doubly-ionized Kr will likely maintain the exhaust-related drift velocity – opening the pathway for phase, multiple-horn, or laser-horn misalignment-based measurements of ion velocity. Digressing, **Figure 6**(d) refers to off-resonant illumination of argon. An increase in $V_B$ generates additional excited species for single-photon ionization.

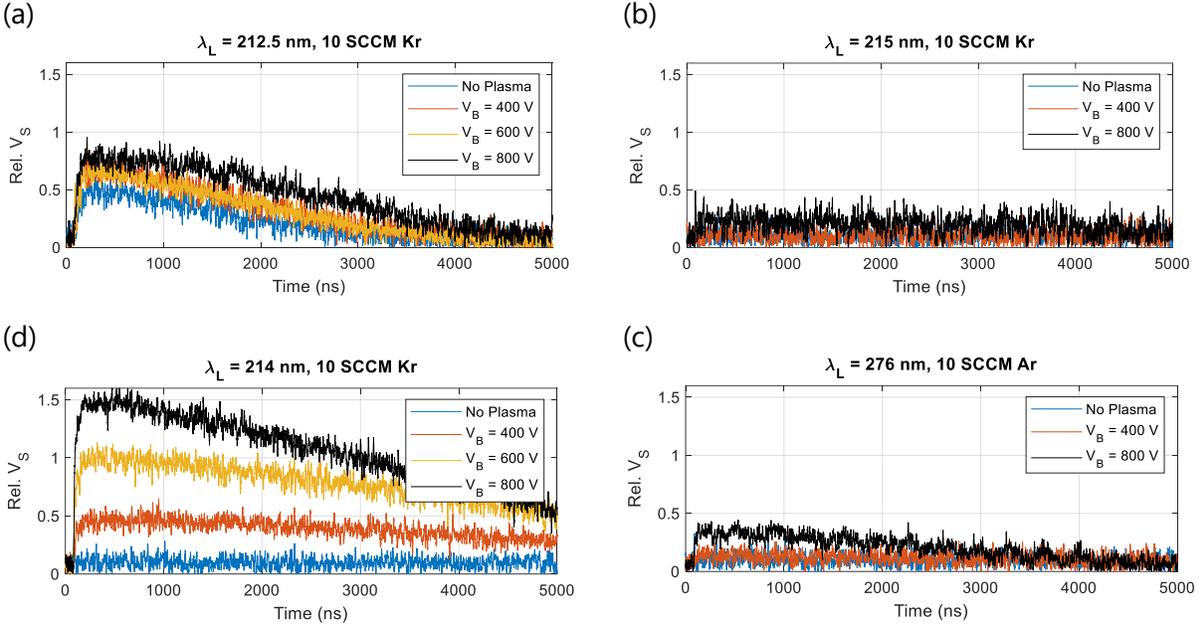

**Figure 6**: Waveforms for constructive elastic microwave scattering-based detection of plasma-embedded laser-induced filaments at several irradiating $\lambda_L$.

The photoionization response (peak $V_S$) can be plotted as a function of $\lambda_L$ (**Figure 7**) to derive information on the underlying plume populations and non-resonant excited-state one-photon PI contribution. Comparison between gaseous and plasma-based krypton emphasizes introduction of the 214 nm resonant-pathway to doubly-ionized Kr III. In the case of argon I, photoionization spectroscopy yields a broad nearly-uniform response until the photon energy (3 eV at 400 nm) becomes comparable with the difference between excited states (11 - 14 eV) and ionization threshold (15.76 eV). Note that resonant photoionization beginning from an excited state (Kr* → Kr* → Kr$^+$) is not observed in **Figure 7**. From energy considerations, these resonances will begin to appear > 500 nm - the subject of future work. However, the technique for measuring relative excited state populations is equivalent to the process depicted in **Table 4** (to be discussed shortly).

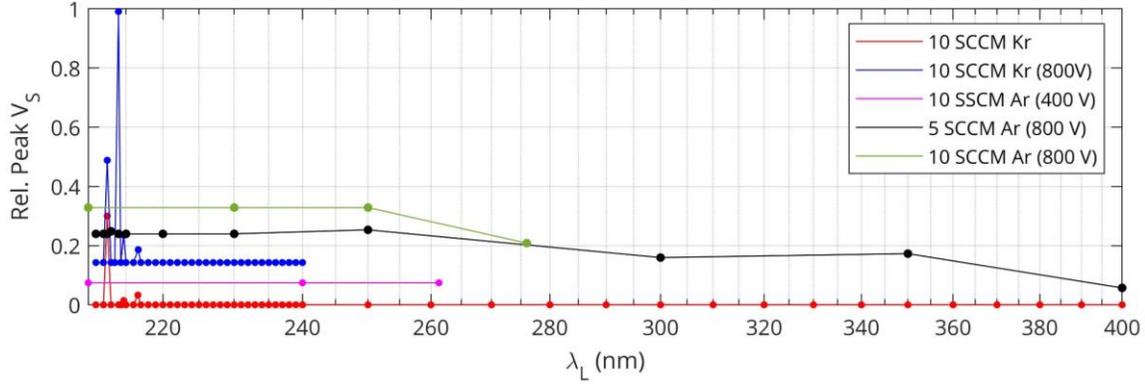

**Figure 7**: CMS+REMPI spectroscopy of underlying plume populations.

A comparison between REMPI-isolated (subtracted non-resonant excited-state one-photon PI contributions from neighboring non-resonant wavelengths), normalized $V_S$ for $\lambda_L = 214$ nm (corresponding to ground state krypton ion populations), normalized Langmuir-probe derived $n_e$, and normalized ion beam-current derived $n_e$ is tabulated in **Table 4**. The case of $\lambda_L = 214.0$ nm bears some-correspondence to the electron number density ($n_e \approx n_i$) and agrees well with Langmuir-based measurements – opening the pathway for normalized $n_e$ mapping (assuming ion populations predominantly lie in the $4p^5\ {}^2P^°_{3/2}$ state).

**Table 4**: Normalized measurements.

|  | Iso. Norm. $V_S$ (214.0 nm) | Norm. LP $n_e$ | Norm. $I_B/\sqrt{V_B}$ |
|---|---|---|---|
| $V_B$ = 400 V, $I_B$ = 14 mA | 0.423 | 0.423 | 0.430 |
| $V_B$ = 600 V, $I_B$ = 30 mA | 0.672 | 0.654 | 0.753 |
| $V_B$ = 800 V, $I_B$ = 46 mA | 1.000 | 1.000 | 1.000 |

## Conclusion

In this work, we reported the first implementation of coherent microwave scattering and multiphoton ionization-based diagnostics in an ion-gridded accelerator. Namely, photoionization spectroscopy of ground and excited-state specie populations – including resonance-enhanced MPI of ground-state ions. Results advocate potential of the technique for relative ion number density ($n_i \approx n_e$) mapping and plume characterization.

# Acknowledgements

This project was supported by the National Science Foundation (Grant No. 1903415). The authors would like to thank Dr. Mikhail Shneider for useful discussions.